# Anderson Localization with Second Quantized Fields: Quantum Statistical Aspects

Clinton Thompson[1], Gautam Vemuri[1*] and G.S. Agarwal[2]

[1]Department of Physics
Indiana University Purdue University Indianapolis (IUPUI)
Indianapolis, IN 46202-3273

[2]Department of Physics
Oklahoma State University
Stillwater, OK 74078

* gvemuri@iupui.edu

**Abstract**: We report a theoretical study of Anderson localization of nonclassical light with emphasis on the quantum statistical aspects of localized light. We demonstrate, from the variance in mean intensity of localized light, as well as site-to-site correlations, that the localized light carries signatures of quantum statistics of input light. For comparison, we also present results for input light with coherent field statistics and thermal field statistics. Our results show that there is an enhancement in fluctuations of localized light due to the medium's disorder. We also find superbunching of the localized light, which may be useful for enhancing the interaction between radiation and matter. Another important consequence of sub-Poissonian statistics of the incoming light is to quench the total fluctuations at the output. Finally, we compare the effects of Gaussian and Rectangular distributions for the disorder, and show that Gaussian disorder accelerates the localization of light.

**References and links**

1. P. W. Anderson, "Absence of Diffusion in Certain Random Lattices," Physical Review **109**, 1492-1505 (1958).

2. See e.g. A. Lagendijk, B. van Tigglen, and D. S. Wiersma, "Fifty Years of Anderson Localization," Physics Today **62**, 24-29 (2009).

3. N. F. Mott, "Metal-Insulator Transition," Reviews of Modern Physics **40**, 677-683 (1968).

4. D. S. Wiersma, P. Bartolini, A. Lagendijk, and R. Righini, "Localization of Light in a Disordered Medium," Nature **390**, 671-673 (1997).

5. S. John, "Localization of Light," Physics Today **44**, 32-40 (1991).

6. See e.g. A. Aspect and M. Inguscio, "Anderson Localization of Ultracold Atoms", Physics Today **62**, 30-35 (2009).

7. L. Sapienza, H. Thyrrestrup, S. Stobbe, P. D. Garacia, S. Smolka, and P. Lodahl, "Cavity Quantum Electrodynamics with Anderson-Localized Modes," Science **327**, 1352 - 1355 (2010).

8. H. B. Perets, Y. Lahini, F. Pozzi, M. Sorel, R. Morandotti, and Y. Silberberg, "Realization of Quantum Walks with Negligible Decoherence in Waveguide Lattices," Physical Review Letters **100**, 170506 (2008).

9. Y. Lahini, A. Avidan, F. Pozzi, M. Sorel, R. Morandotti, D. N. Christodoulides, and Y. Silberberg, "Anderson Localization and Nonlinearity in One-Dimensional Disordered Photonic Lattices," Physical Review Letters **100**, 013906 (2008).

10. A. Rai, J. Das, A.S. Agerwal, "Quantum Entanglement in Coupled Lossy Waveguides," Optics Express **18**, 6241-6254 (2010).

11. S. Longhi, "Optical Analog of Population Trapping in the Continuum: Classical and Quantum Interference Effects," Phys. Rev. A **79**, 023811 (2009).

12. Y. Bromberg, Y. Lahini, R. Morandotti and Y. Silberberg, "Quantum and Classical Correlations in Waveguides Lattices", Phys. Rev. Lett. **102**, 253904 (2009).

13. S. Longhi, "Optical Bloch Oscillations and Zener Tunneling with Nonclassical Light", Phys. Rev. Lett. **101**, 193902 (2008).

14. A. Rai, G.S. Agarwal and J.H.H. Perk, "Transport and Quantum Walk of Nonclassical Light Coupled Waveguides", Phys. Rev. **A78**, 042304 (2008).

15. A. Politi, M.J. Cryan, J.G. Rarity, S. Yu and J.L. O'Brien, "Silica-on-Silicon Waveguide Quantum Circuits", Science **320**, 646-649 (2008).

16. S. Longhi, "Transfer of Light Waves in Optical Waveguides via a Continuum", Phys. Rev. **A78**, 013815 (2008).

17. 17. Y. Lahini, Y. Bromberg, D. N. Christodoulides, and Y. Silberberg, "Quantum Correlations in Anderson Localization of Indistinguishable Particles," http://arxiv.org/abs/1003.3657v1

18. M. O. Scully, and M. S. Zubairy, *Quantum Optics* (Cambridge University Press, Cambridge, New York, 1997).

19. W. H.Press, S. A. Teukolsky, W. T. Vetterling, and B. P. Flannery, *Numerical Recipes in C++: The Art of Scientific Computing* (Cambridge University Press, Cambridge, New York, 2002).

20. B.R. Mollow, "Two Photon Absorption and Field Correlation Functions", Physical Review **175**, 1555-1563 (1968).

**Introduction**

The localization of the wavefunction of a particle, due to the disorder of the medium through which it propagates, has been extensively studied since its first examination by Anderson.[1] Anderson localization (AL) has, since then, been explained as the interference between the probability amplitudes of the different, and competing, pathways that the particle can take within the medium, and in terms of cancellation of intensities in all directions but the one along which the localization occurs.[2] The most remarkable aspect of AL is the possibility of the emergence of coherence in a strongly disordered medium. While the original studies of AL were in the context of electrons propagating through various disordered media, such as semiconductors,[3] later studies considered the localization of light in random media,[4] and photonic waveguides.[5] There is extensive literature on Anderson localization of matter waves using ultracold atoms.[6] Some important applications of AL have started appearing. For example, very recently, AL has been used to enhance the coupling between photons and atoms by localizing the light that interacts with the medium in a cavity QED experiment.[7]

This manuscript reports on a theoretical study of light propagating through a medium which consists of an array of evanescently coupled waveguides.[8,9] These coupled waveguides are excellent systems for studying AL because one can introduce well defined disorder and thus fine features of AL can be studied in such optical systems. Further, due to advancements in the production of nonclassical light sources, we can study AL with second quantized fields. In particular, our interest is in investigating the nature of quantum statistics of AL when the input light has different photon statistics, with special emphasis on squeezed light. This is an issue that can be studied in detail using various types of available light sources. Quantum statistical features can be probed by studying fluctuations in the expectation value of a desired observable. In our case, we study the variance in the intensity of localized light. We thus pay special attention to the function $g^{(2)}$ introduced by Glauber to characterize quantum fields. To put the results for squeezed light in perspective, we also present some results for AL with classical coherent light and classical thermal light. Note that recently several papers have reported on the propagation of nonclassical light in waveguides without disorder.[10-16] Finally, we also study the consequences of having a Gaussian versus a Rectangular distribution for the disorder of the medium.

The principal results of our study are that when the medium is assumed to have a Gaussian disorder, there is an enhancement of the disorder's effect on AL relative to a Rectangular distribution for the disorder. We also find that the variance in the intensity fluctuations, at the waveguide into which light is localized, increases with disorder, with the magnitude of the fluctuations being largest for squeezed light and least for coherent light. The site-to-site correlations show that for localized light, there is a superbunching of light into the waveguide in which the light is initially coupled. Finally, we show analytically that for two photons coupled into the medium, there is a suppression of the fluctuations by the disorder of the medium.

A recent manuscript has also investigated some quantum statistical aspects of AL.[17] However, the emphasis in that manuscript is on very different features. Specifically, that manuscript studies the evolution of particles in a disordered medium, where the particles obey either Bose-Einstein statistics or Fermi-Dirac statistics. The authors find evidence for bunching and antibunching, depending on the statistics of the particles, and on the disorder of the medium and the distance within the waveguide over which the evolution takes place.

**MODEL**

Consider an array of single mode waveguides with neighboring waveguides coupled evanescently. Since we are specifically interested in second quantized fields we describe fields in each waveguide by the Bosonic operators $a_j$ and $a_j^+$. These field operators obey the commutation relations

$$[a_j, a_j^+] = 1, \quad [a_j, a_j] = 0, \quad [a_j^+, a_j^+] = 0 \qquad (1)$$

The Hamiltonian describing propagation in a waveguide array would be

$$H = -C[\sum_j a_{j+1}^+ a_j + h.c.] + \sum_j \beta_j a_j^+ a_j \qquad (2)$$

where the evanescent coupling between neighboring waveguides is denoted by C and where $\beta_j$ is an effective detuning parameter for the jth waveguide. Using Eq. (2), we write Heisenberg equations for the field operators

$$i\frac{\partial a_j}{\partial z} = C(a_{j+1} + a_{j-1}) + \beta_j a_j \qquad (3)$$

In order to study Anderson localization of photons one can now introduce disorder either through the coupling constant C or through $\beta_j$. We adopt the latter. Thus $\beta_j$ are taken to be real and random. Specifically, the deviation in $\beta_j$ from its mean is assumed to be a random variable. The mean value of $\beta_j$ is irrelevant as long as it is the same for all waveguides. We can thus set it to be zero. We further assume that the random variables $\beta_j$ are independent of each other. In this paper, we consider two types of distributions for the disorder of the medium – Gaussian and Rectangular. The Gaussian distribution is given by

$$P(\beta) = \frac{1}{\sqrt{2\pi\Delta^2}} \exp(\frac{-\beta^2}{2\Delta^2})$$ where $\Delta^2$ is the variance of the distribution and is a measure of the disorder in the medium. The Rectangular distribution has a probability distribution of the form $P(\beta) = \frac{1}{2\Delta}$ for $|\beta| \leq \Delta$ and zero otherwise. Now, in view of the linearity of the Heisenberg equations, we can write the solution to Eq. (3) as

$$a_l = \sum_{l'} G_{ll'} a_{l'}(0) \qquad (4)$$

where $a_{l'}(0)$'s are the Heisenberg operators at the input port of the waveguides. Note that the Green's function G depends on the parameters C and $\beta_j$ and is random in nature due to disorder in $\beta_j$. All the physical quantities at the output would require averaging of the Greens function and its powers. The quantized nature of the fields enters through the input Heisenberg operators.

Now we discuss what could be measurable quantities. Clearly mean intensities, $I_l$, at the output are obvious measurable quantities and these are given by

$$I_l = \langle a_l^+ a_l \rangle = \sum_p \sum_q < G_{l,p}^* G_{l,q} >< a_p^+(0) a_q(0) > \qquad (5)$$

where the product of Green's functions is to be averaged over the ensemble of distributions of $\beta_j$. In this paper we focus on the input light in a single waveguide, labeled as zero-then Eq. (5) simplifies to

$$I_l = < G_{l,0}^* G_{l,0} >< a_0^+ a_0 > \qquad (6)$$

In order to examine the quantum statistical aspects of localization we can study the fluctuations in the intensity at the output. Glauber introduced the function $g^{(2)}$ defined by

$$g^{(2)} = <a^{+2}a^2> / <a^+a>^2 \qquad (7)$$

Note that values of $g^{(2)}$ greater [smaller] than one correspond to bunching [antibunching]. Using the solution in Eq. (4), $g^{(2)}$ can be written in terms of the Greens function as

$$g^{(2)} = \frac{<|G_{l,0}|^4><a_0^{+2}a_0^2>}{<G_{l,0}^*G_{l,0}>^2<a_0^+a_0>^2} \qquad (8)$$

Note that the quantum statistical quantity $g^{(2)}$ involves the averages of fourth powers of the Greens function. It is at this point that we start getting newer aspects of Anderson localization with quantized fields. All previous works, except Ref. 17, essentially correspond to the study of the quantity appearing in the equation for mean intensity, i.e. Eq. (6).

To further probe the effect of input light statistics on the quantum statistical aspects of AL, we calculate site-to-site correlations defined by

$$\langle I_l I_p \rangle = \left\langle |G_{l0}|^2 |G_{p0}|^2 \right\rangle <a_0^{+2}a_0^2>. \qquad (9)$$

The physical quantities introduced above do require the nature of the input fields. We will consider three types of input fields – (i) a coherent field, i.e. a field in a coherent state $\alpha_o$, (ii) a thermal field with average photon number $n_0$, and (iii) a nonclassical field, such as a field in a squeezed state.[18] For all these fields, the quantities that we need in the above calculations are given as follows:

for a coherent field, $\left\langle a_o^{+2} a_o^2 \right\rangle = |\alpha_o|^4$ and $\left\langle a_o^+ a_o \right\rangle = |\alpha_o|^2$, (10a)

for a thermal field, $\langle a_o^{+2} a_o^2 \rangle = 2n_0^2$ and $\langle a_o^+ a_o \rangle = n_0$, (10b)

and for single-mode, squeezed field

$$\langle a_o^{+2} a_o^2 \rangle = \sinh^2 r (1 + 3\sinh^2 r) \text{ and } \langle a_o^+ a_o \rangle = \sinh^2 r. \qquad (10c)$$

In Eq. (10c), r is the squeezing parameter. When we compare final results for different input fields we will assume that all fields have the same average photon number, i.e.

$$n_0 = |\alpha_o|^2 = \sinh^2 r$$

For the numerical results below, we assume that the medium consists of 100 waveguides, and that the input light is coupled into the 50[th] waveguide. The random disorder is generated as follows: computer generated, uniformly distributed random numbers are used to generate 100 Gaussian distributed random numbers with zero mean and desired variance via the Box-Mueller algorithm.[19] Eq. (3) is solved numerically, with each value of $\beta_j$ corresponding to one waveguide. The output intensity is computed from the expression $\langle a_j^+ a_j \rangle$ (=I) where the angular brackets represent averaging over 1000 realizations of the disorder and the quantum mechanical average over the input fields.

**Results**

This section describes the results of our studies on the quantum statistical aspects of AL for different input photon statistics. For numerical calculations, we take C = 1, and the average value of $\beta$ as zero. We begin with the injection of a classical coherent field, a thermal field and a squeezed field at waveguide 50. Figure 1(a) – 1(c) show the mean intensity as a function of the waveguide position for all three input photon statistics. Note that each figure has three light distributions, but that the distributions for the three different field statistics are indistinguishable. We assume that the mean number of photons in the input field is 100, and that the disorder of the medium has a Gaussian distribution. Clearly, as the disorder is increased, the output field is localized, until at a disorder of about $\Delta/C = 3$, all three output light patterns converge to the same distribution. Note that after the localization has taken place, about half of the input photons are found at the output of the waveguide through which the input fields were sent. This characteristic property is essentially the reason for the enhanced radiation-matter interaction using localized modes.[6]

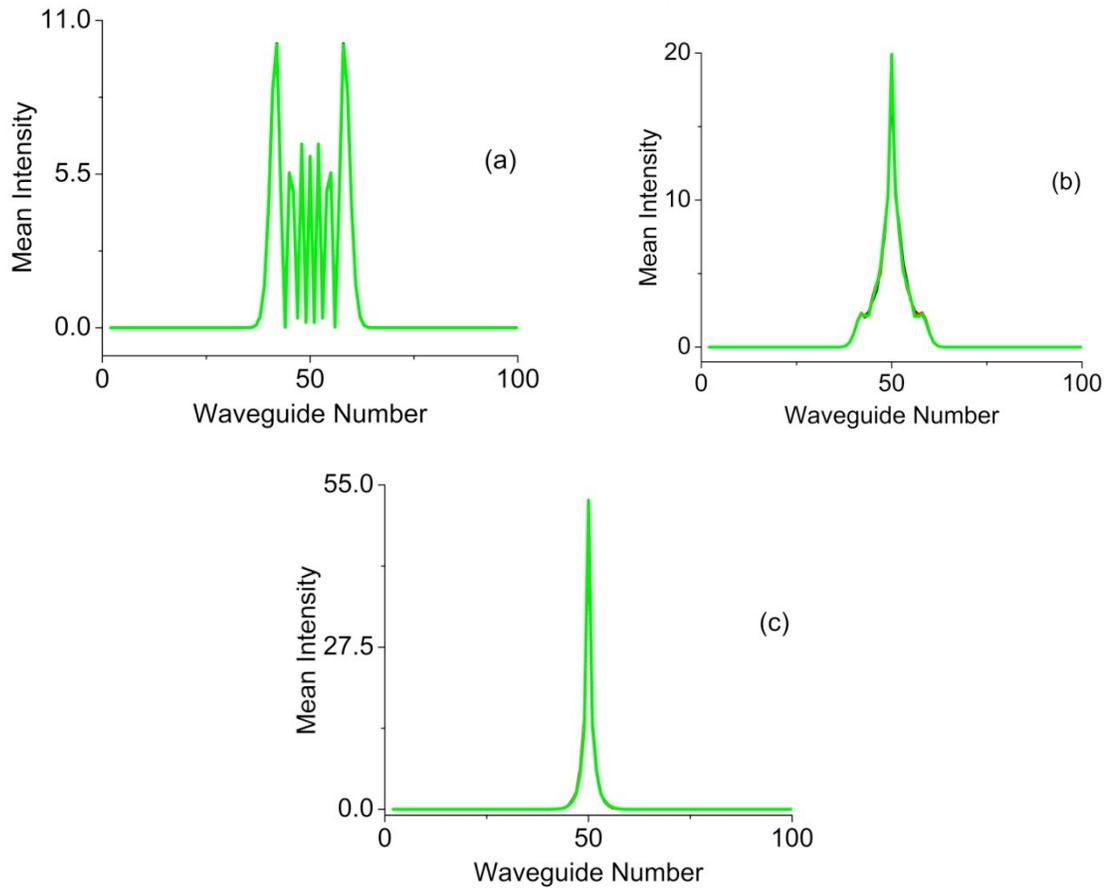

*Figure 1: Mean intensity vs. waveguide number for disorders of (a) $\Delta/C = 0$, (b) $\Delta/C = 1$ & (c) $\Delta/C = 3$. Each plot shows three indistinguishable curves for the three different input*

*photon statistics (coherent, thermal and squeezed). Mean photon number for all three input fields is 100.*

Figure 2 shows the evolution of the light patterns for Gaussian and Rectangular distributions of the disorder. Clearly, for a given amount of disorder, a Gaussian distribution for the disorder leads to greater localization of light, suggesting that a Gaussian distribution enhances the effect of the medium's disorder.

This enhancement of the disorder by a Gaussian distribution can be understood as follows. The variance in $\beta_j$ for the Gaussian distribution is given by $\Delta^2$ as seen from the form of the distribution given earlier. The variance for the Rectangular distribution is given $\Delta^2/3$. Thus, for a given $\Delta$, a Gaussian disorder effectively increases the disorder in the medium as compared to a Rectangular distribution.

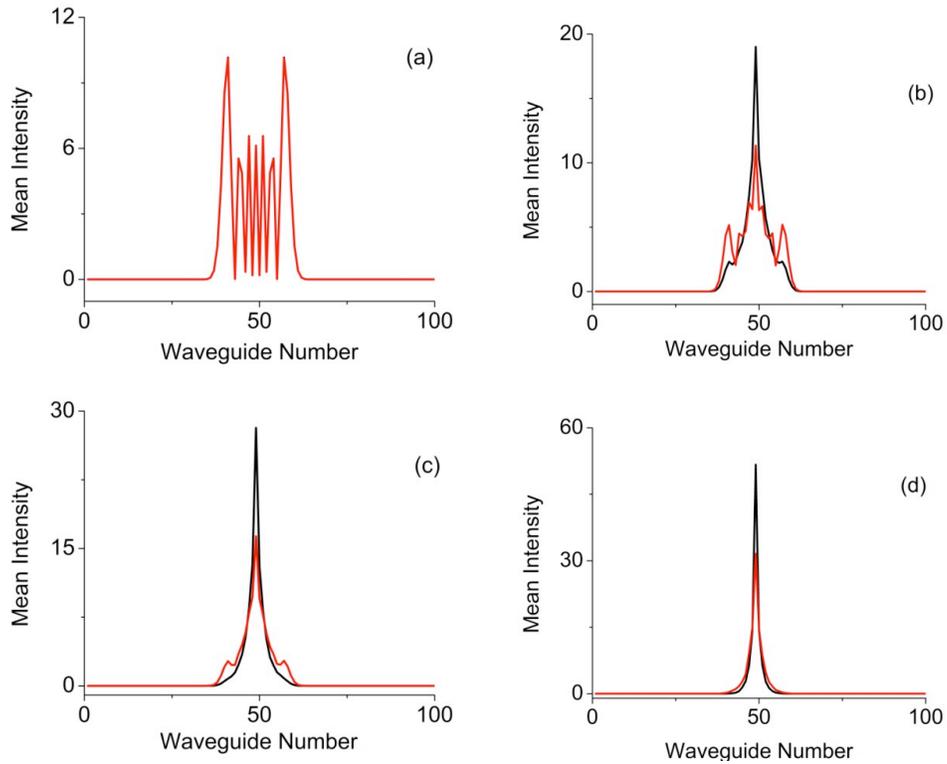

*Figure 2: Mean intensity vs. waveguide number for disorders of (a) Δ/C = 0, (b) Δ/C = 1, (c) Δ/C = 1.5 & (d) Δ/C = 3. Each plot has two curves, one for the Gaussian disorder (black) and one for a Rectangular disorder (red). Mean photon number for all input fields is 100.*

For the localized light, the mean number of output photons for a rectangular distribution for the medium disorder is less than that for Gaussian distribution and hence for applications of the type utilized in Ref. 7, it may be preferable to use Gaussian distributions.

Next, we present results for the variance in the mean intensity. Figure 3(a) shows the variance in the mean intensity of the output light at the $50^{th}$ waveguide for a Gaussian distribution for the disorder. It is seen that the variance in the intensity increases with disorder, which is somewhat counter-intuitive since one might expect the variance to reduce at localization. The variance is the largest for squeezed light, and least for a coherent field. Figure 3(b) shows the variance for a Rectangular distribution for the disorder, and the qualitative trends are similar to those for the Gaussian disorder. However, the magnitude of the variance for a given disorder is less for Rectangular disorder than for Gaussian disorder, which again implies that a Gaussian disorder enhances the intensity fluctuations in localized light more than a Rectangular distribution. Note that the magnitude of the variance in Fig. 3 is sensitive to the mean number of input photons; however, the qualitative trends in the variance are similar.

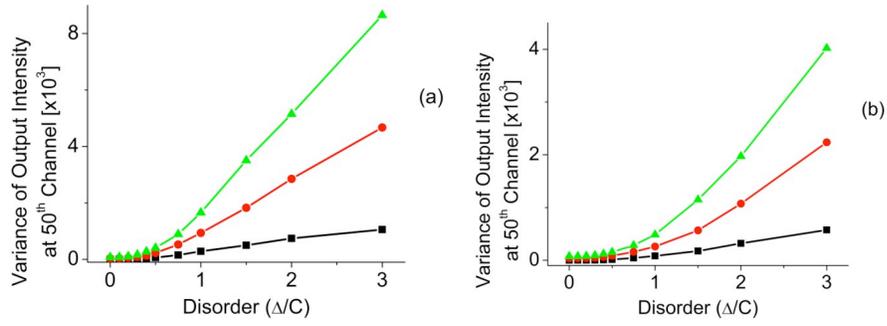

*Figure 3: Variance at the output of the 50$^{th}$ waveguide vs. disorder for (a) Gaussian disorder & (b) Rectangular disorder. Mean photon number for all three input fields is 100. Data shown are for input photon statistics of coherent field (black), thermal field (red) and single-mode, squeezed field (green).*

To gain some additional insight on the fluctuation behavior of AL, we show the quantity, $g^{(2)}$ (ratio of variance to square of the mean) at the 50$^{th}$ waveguide for different input photon statistics in Fig. 4. The mean photon number for all three inputs is 100 and the disorder is taken to be a Gaussian [Rectangular] distribution in Fig. 4(a) [Fig. 4(b)]. An interesting feature here is the enhancement of $g^{(2)}$ by the disorder of the medium.

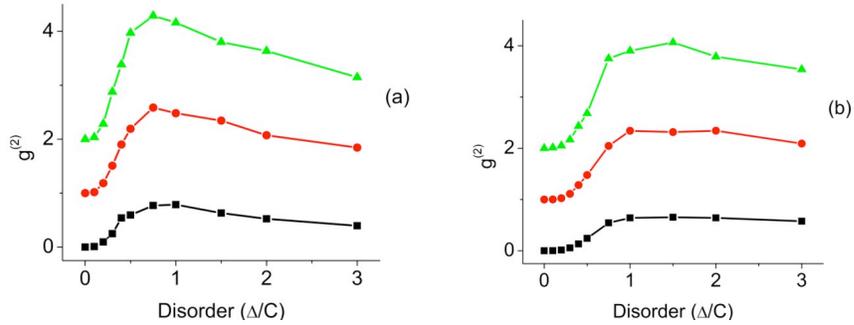

*Figure 4: Normalized variance vs. disorder at 50$^{th}$ waveguide for (a) Gaussian disorder & (b) Rectangular disorder. Mean photon number for all three input fields is 100. Curves shown are for coherent fields (black), thermal fields (red) and squeezed fields (green).*

Let us consider the case of single-mode, squeezed light at the input. The normalized variance for the input light is 2, whereas it increases to more than 4 for a disorder of around 1. For higher disorders, i.e. after complete localization, $g^{(2)}$ is still higher than for zero disorder. A similar enhancement of fluctuations is seen with thermal light, and to a smaller extent with coherent light input. Very similar behavior is observed for a Rectangular distribution for the disorder, with the exception that the maximum in $g^{(2)}$ occur for higher disorders, and the enhancement is less than for Gaussian disorder.

We now turn to the details of the case where the input light is a single-mode, squeezed field. Fig. 5 is the variance in the mean intensity at the $50^{th}$ waveguide as a function of the squeezing parameter for a disorder, $\Delta/C \sim 3$. For small values of r, a situation that is similar to having two photons, the variance is very small. However, with an increase in r, there is a rapid increase in the variance at the waveguide at which the AL occurs.

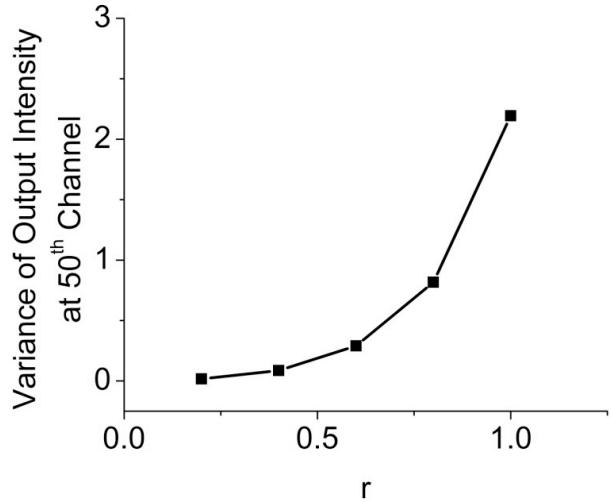

*Figure 5: Variance in output intensity at $50^{th}$ waveguide vs. squeezing parameter, for a Gaussian disorder.*

The next observable we present is the site-to-site correlations, i.e. the quantity, $\langle I_l I_p \rangle = \langle |G_{10}|^2 |G_{p0}|^2 \rangle < a_0^{+2} a_0^2 >$, discussed earlier, for the case where the input light is a single-mode, squeezed field. For small r and no disorder in the medium, Fig. 6(a), we find that the magnitudes of the correlations are small, and it is apparent that in the absence of localization, there is a near equal probability for the output photons to be in waveguides

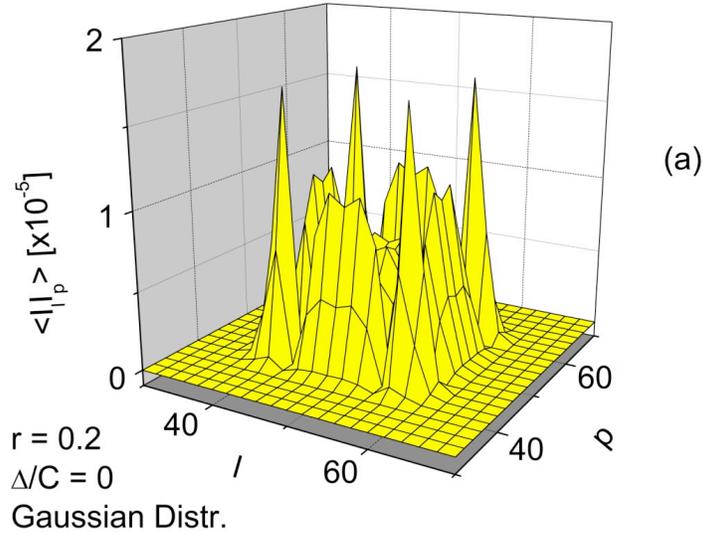

*Figure 6(a): Site-to-site correlation functions for r = 0.2 and Δ/C = 0*

from 40 to 60. With an increase in the disorder (see Fig. 6b & 6c), there is a superbunching of the photons into the waveguide into which the input photons were launched, and a diminishing probability for the photons to be found in adjacent waveguides. Of course, the magnitudes of the correlations are still small, due to the small value of the squeezing parameter. It is also seen that the spread in the photon distribution is greater for a Rectangular disorder than for a Gaussian disorder, consistent with the earlier observation that a Gaussian disorder leads to greater localization of light. In Fig 6(d) and 6(e) are the shown the site-to-site correlations for r = 1 and a disorder, Δ/C = 3. The evidence for superbunching of the output photons is quite pronounced, and there is negligible probability of the photons spreading more than about 5 waveguides on either side of the 50$^{th}$ waveguide. This superbunching may have possible utility in enhancing nonlinearities in the interactions between radiation and matter since we know from Mollow's work [20] that two-photon absorption in a thermal field is enhanced over that that in a coherent field.

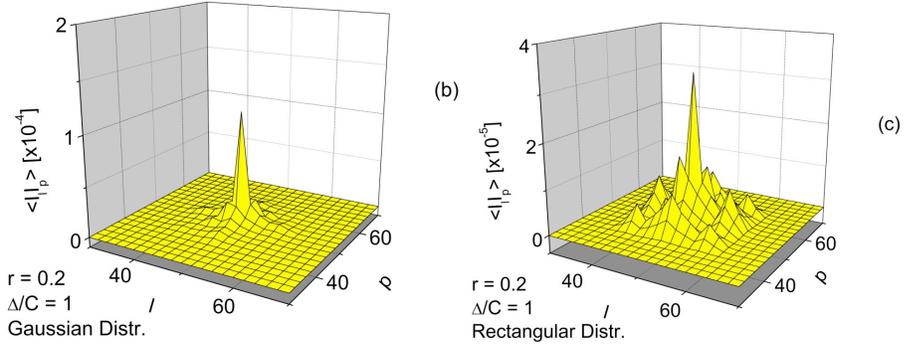

*Figure 6(b) & 6(c): Site-to-site correlation functions for r = 0.2 and Δ/C = 1.*

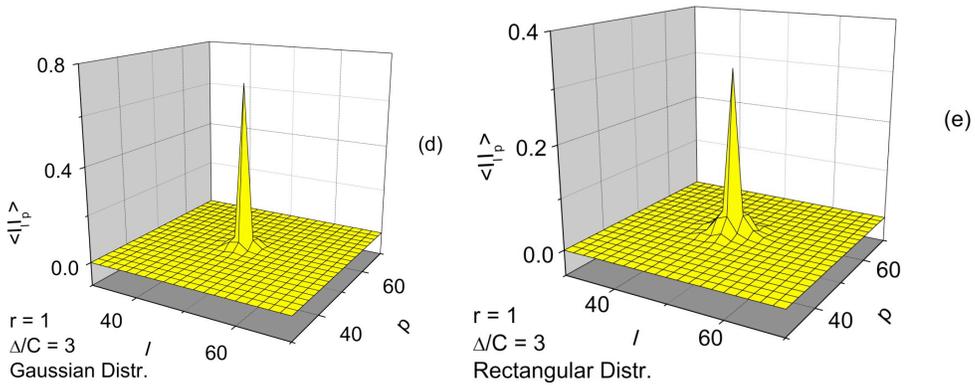

*Figure 6(d) & 6(e): Site-to-site correlations for r = 1 and Δ/C = 3.*

Let us finally consider the case when two photons are launched into the disordered medium. Specifically, we take a Fock state with $<a_0^{+2} a_0^2> = 2$ and $<a_0^+ a_0> = 2$. Since $g^{(2)}$ is given by

$$g^{(2)} = \frac{<|G_{l,0}|^4>}{<|G_{l,0}|^2>^2} \cdot \frac{<a_0^{+2} a_0^2>}{<a_0^+ a_0>^2} \qquad (11)$$

one can see that the first ratio (containing the Greens functions) provides information on the fluctuations induced by the disorder in the medium, whereas the second ratio is specific to the statistics of the input photons. In particular, this ratio is equal to 1 for a coherent state and ½ for a Fock state, which immediately suggests that the output field is less noisy in the latter case. This result demonstrates an instance in which there is a suppression of the fluctuations due to the disorder of the medium by the nonclassical sub-Poissonian statistics of the input field.

**Conclusions**

This manuscript has focused on the quantum statistical aspects of localized light when the input light has photons statistics of coherent light, thermal light or single-mode, squeezed light. By numerically solving the Heisenberg equation for the field operators, we have calculated relevant quantum statistical observables, such as the variance in the intensity fluctuations of localized light, site-to-site correlations and the Glauber $g^{(2)}$ function. We have also reported on a comparison of the effect of the statistics associated with the disorder of the medium on Anderson localization and the associated quantum statistics.

Our study shows that a Gaussian distribution for the disorder enhances the effect of the medium's disorder on AL when compared to a Rectangular distribution. Furthermore, a Gaussian distribution enhances the fluctuations in the intensity of the localized light to a greater extent. By calculating the variance in the intensity fluctuations at the waveguide into which light is localized, we have shown that the fluctuations increase with disorder. The $g^{(2)}$ function has a maximum for a finite disorder, before it tapers off for higher disorders to a value that is still greater than that for zero disorder.

The site-to-site correlations show that the probability of finding photons in waveguides that are adjacent to the one into which the input light is coupled diminishes with increasing disorder. For sufficiently high disorder, we find a superbunching of light into the waveguide in which localization occurs.

Finally, we have shown analytically that there are some instances in which there is a suppression of fluctuations by the disorder of the medium and nonclassical sub-Poisonnian statistics of the input light.

**Acknowledgments**

C.T. was supported by a fellowship from a GAANN award from the U.S. Department of Education to GV.